\documentclass{webofc}
\usepackage[varg]{txfonts}   
\newcommand{\GeV}{\ensuremath{\mathrm{GeV}}}
\newcommand{\mus}{\ensuremath{\mu}s\xspace}
\newcommand{\ttbar}{\ensuremath{\mathrm{t}\overline{\mathrm{t}}}}

\def\pt{\ensuremath{p_{\mathrm{T}}}\xspace}

\usepackage{algorithmic}
\usepackage{xspace}
\usepackage{xfrac}
\usepackage{units}
\usepackage{lineno}
\usepackage[linktocpage]{hyperref}
\hypersetup{breaklinks=false}
\usepackage[all]{hypcap}
\usepackage[font=small,labelfont=bf,tableposition=top]{caption}

\begin{document}
\title{Level-1 Track Finding with an all-FPGA system at CMS for the HL-LHC}
%
%

\author{\firstname{Zhengcheng} \lastname{Tao}\inst{1}\fnsep\thanks{\email{zt75@cornell.edu}}, 
        \lastname{for the CMS Tracker Group}
        }

\institute{Cornell University, Ithaca, NY, USA}

\abstract{%
  With the High Luminosity LHC upgrades, incorporating tracking information into the CMS Level-1 trigger becomes necessary in order to maintain a manageable trigger rate and good trigger performance e.g. to retain thresholds for electroweak physics. The main challenges Level-1 track finding faces are the large data throughput from the detector at a collision rate of 40 MHz and a 4 \mus latency budget to reconstruct charged particle tracks with sufficiently low transverse momentum to be used in the Level-1 trigger decision. Dedicated all-FPGA hardware systems with time-multiplexed architecture have been developed for track finding to address these challenges. The algorithm and performance of the pattern recognition and particle trajectory determination are discussed. The implementation on customized electronics with commercially available FPGAs is presented as well.
}
\maketitle
\section{Introduction}
\label{sec:intro}
The Large Hadron Collider (LHC) at CERN is scheduled to undergo major upgrades during 2024 - 2026. The upgraded collider complex, High-Luminosity LHC (HL-LHC), is expected to deliver an instantaneous peak luminosity up to $\unit[7.5 \times10^{34}]{\mathrm{cm}^{-2}\mathrm{s}^{-1}}$ at a center-of-mass energy of 14 TeV~\cite{hllhc}. 
This corresponds to an average number of overlapping proton-proton collisions, named pileup (PU), up to 200 per bunch crossing at 40 MHz. 
While the HL-LHC will provide great opportunities for e.g. precision Higgs measurements and searches for exotic processes with small cross sections, it is also very challenging to build a detector that can fully exploit the physics potential delivered by the HL-LHC, especially for the trigger systems.

The CMS experiment~\cite{cms} adopts a two-level trigger system~\cite{cms_trigger} to select data that are of interest for further studies: the real-time hardware-based Level-1 (L1) trigger and the software-based High Level Trigger (HLT). 
The current CMS L1 trigger uses only information from the calorimeters and the muon system to make L1 trigger decisions. Under the HL-LHC pileup condition, the trigger rate with the current L1 trigger would exceed the capability of the front-end electronics. 
Significantly increasing the trigger threshold with the hope of reducing the trigger rate not only would alone be insufficient, but also limit the physics potential by lowering the efficiency for the physics of interest. 
It is necessary to integrate the charged particle tracking information into the current L1 trigger system, adding extra handles to manage the trigger rate. Using tracking information in the L1 trigger would improve identification and transverse momentum (\pt) resolution of charged particles. It also provides additional vertex and track isolation information for triggering on hadronic activities. 

\begin{figure}
\begin{center}
\includegraphics[width=6.25cm]{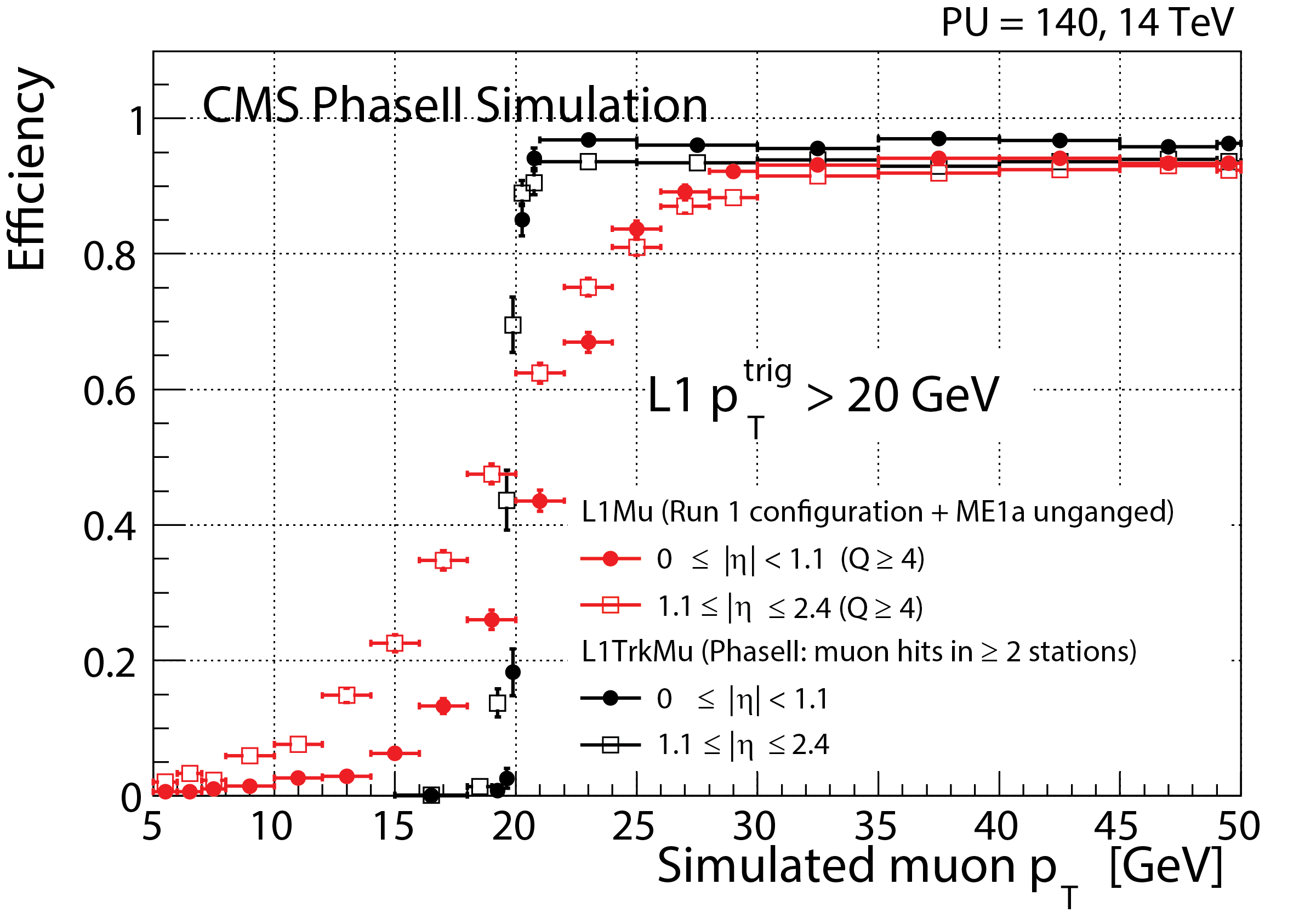}
\includegraphics[width=6.25cm]{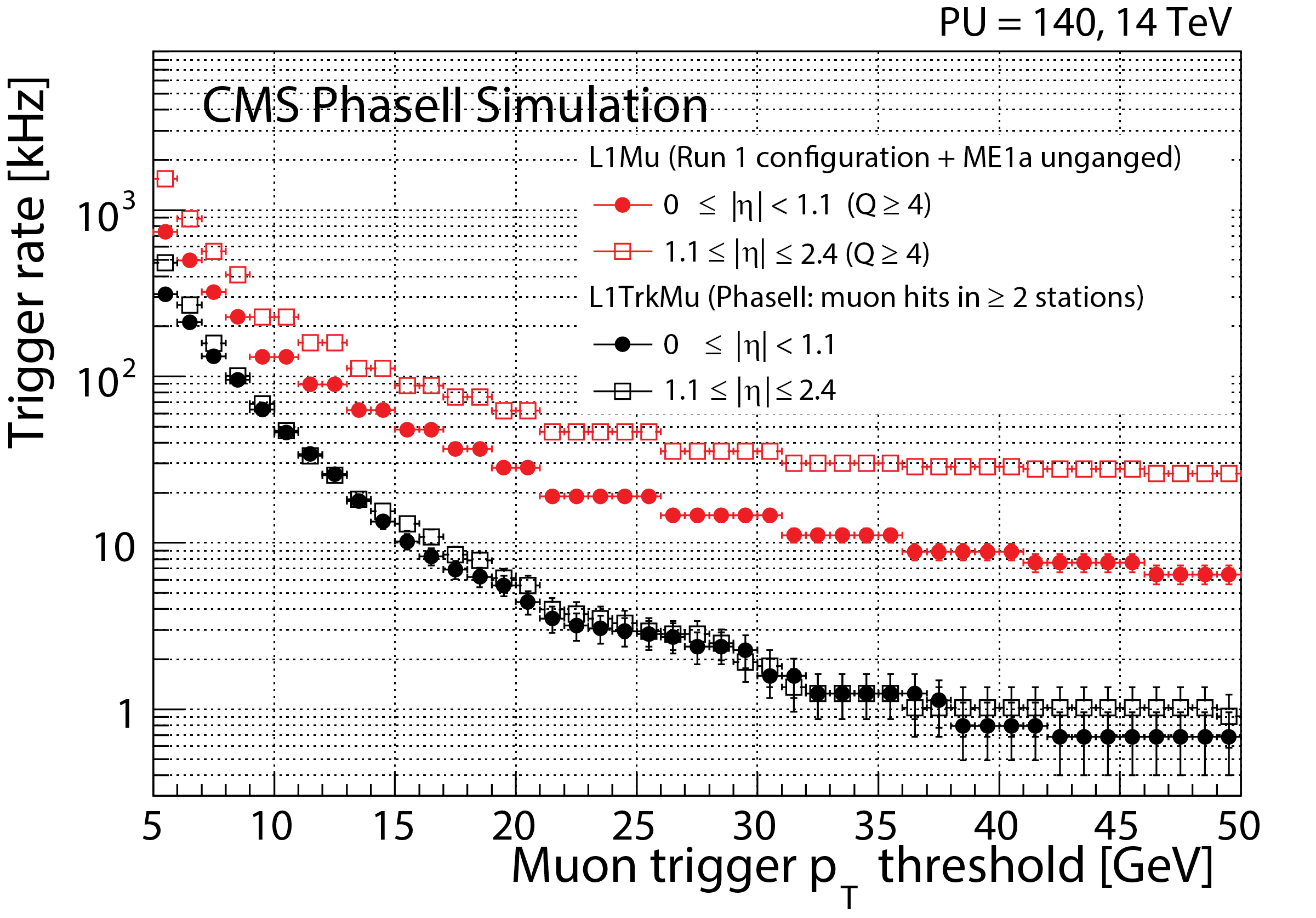}
\caption{ Single muon trigger efficiency with $\pt> \unit[20]{GeV}$
  as a function of muon \pt (left) and the trigger rate as a function of the
  muon trigger threshold (right). Both the stand-alone muon
  trigger (red) and when including L1 tracking (black) are shown for 
  $|\eta|<1.1$ (solid circles) and $1.1 \leq |\eta| \leq 2.4$ (hollow circles)~\cite{tp}.}
\label{fig:trigeff}
\end{center}
\end{figure}

Figure~\ref{fig:trigeff} shows an example of the L1 single muon trigger performance with PU = 140~\cite{tp}. The plot on the left shows the trigger efficiency with \pt threshold of 20 GeV as a function of simulated muon \pt. The plot on the right shows the trigger rate as a function of trigger \pt threshold. The red dots correspond to the stand-alone L1 muon trigger, while the black dots represent the same trigger but incorporating L1 tracking information. 
Without tracking at L1, it is more difficult for the stand-alone muon trigger to determine the muon momentum at higher \pt. Therefore, increasing the trigger \pt threshold becomes less effective to reduce the trigger rate. Adding tracking information at L1 provides a better \pt measurement, leading to a more effective reduction of the trigger rate, as well as a sharper turn-on at a given trigger threshold.

Two approaches to implement the track finding at the L1 trigger are presented in this proceeding. Both approaches are based on the commercially available Field Programmable Gate Array (FPGA) technology. 
The algorithms are discussed in Sect.~\ref{sec:algorithm}, following a brief overview of the CMS tracker upgrades for the HL-LHC (Phase-2 upgrades) in Sect.~\ref{sec:tracker}. The implementations of these algorithms on the hardware demonstrators are presented in Sect.~\ref{sec:demo}. The performance of the two L1 track finding approaches is shown in Sect.~\ref{sec:performance}.

\section{The CMS Tracker Upgrades}
\label{sec:tracker}
\begin{figure*}
\begin{center}
\includegraphics[width=0.8\linewidth]{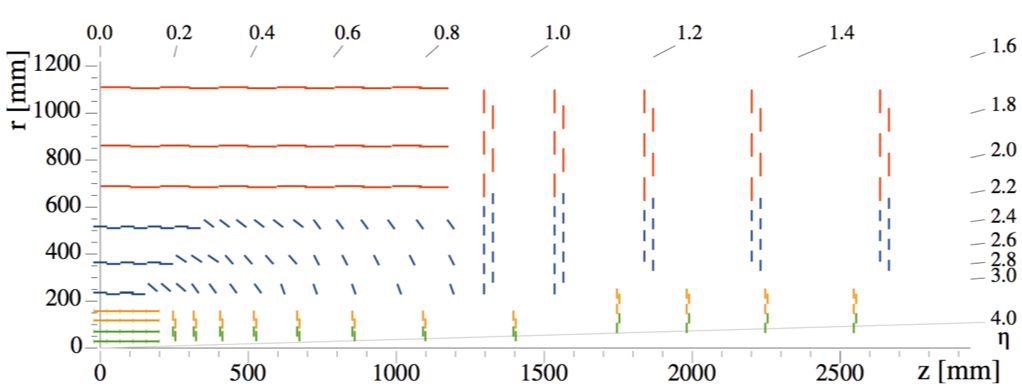}
\caption{One quarter of the proposed CMS Phase-2 tracker layout in the $r-z$ plane, 
showing the placement of both PS (blue) and 2S (red) modules. 
The orange and green modules are of the inner pixel tracker and are not used in the L1 track finding~\cite{tracker_tdr}.}
\label{fig:phase2tracker}
\end{center}
\end{figure*}

In preparation for the HL-LHC era, the current CMS silicon tracker will be completely replaced during Long-Shutdown 3 (LS3) before the operation of the HL-LHC.
Figure~\ref{fig:phase2tracker} shows a quarter of the proposed tracker layout in the $r-z$ plane.
The new CMS tracker will be designed to allow, for the first time, the incorporation of tracking information into the L1 trigger system.
To achieve this, local data reduction in the front-end readout electronics is necessary in order to limit the data volume that is sent out at the 40 MHz collision rate.
This is achieved with a novel module design, referred to as ``\pt module'', that is able to reject signals from particles below a certain \pt threshold~\cite{tracker_tdr}. 
The \pt module is composed of two single-sided closely-spaced sensors, read out by a common set of front-end electronics. The \pt modules are able to provide \pt discriminations by correlating the signals in the two sensors and sending out only the hit pairs (``stubs'') compatible with particle tracks above the chosen \pt threshold based on the bend of the hit pairs in the CMS magnetic field.
Two different types of \pt modules are employed: pixel-strip (PS) modules and strip-strip (2S) modules. 
The PS modules are capable of providing precise $z$ or $r$ coordinate measurements, and are used in the first three barrel layers of the outer tracker as well as in the endcaps with radii less than about 700 mm.
The 2S modules are deployed in the outermost three barrel layers and in the endcaps at larger radii.
Stubs from the \pt modules are the inputs to the L1 track finding system.

\section{L1 Track Finding Approaches}
\label{sec:algorithm}
The goal of the L1 track finding is to reconstruct charged particle trajectories with $\pt>2~\GeV$. The system needs to be able to handle high data throughput from the detector at 40~MHz and reconstruct tracks within about 4 \mus in order to be used in the L1 trigger decisions.
The two approaches presented in this proceeding, referred to as ``Tracklet''~\cite{jorge, louise, margaret} and ``TMTT''~\cite{tmttelba, tmtt_demo_paper} respectively, address the above challenges via highly parallelized algorithms. The parallelizations are achieved both in space, by partitioning the detector into smaller regions, and in time, by using time-multiplexed architecture. For a fixed time-multiplexing factor $n$ (typically 6--18), each time slice of the system processes a new event every $n \times 25~\mathrm{ns}$, at a cost of making $n$ duplicates of the system.
The two approaches, though different in details, consist of the following major steps: data organization, pattern recognition, track fitting and duplicate removal. 
Both approaches are pipelined designs with fixed latencies, and are based on all-FPGA systems. 
The ever-increasing capability and programming flexibility of commercially available FPGAs make them ideal for the task of fast track finding. 

\subsection{Tracklet algorithm}
\label{sec:tracklet_algo}
\begin{figure*}
\begin{center}
\includegraphics[width=0.3\linewidth]{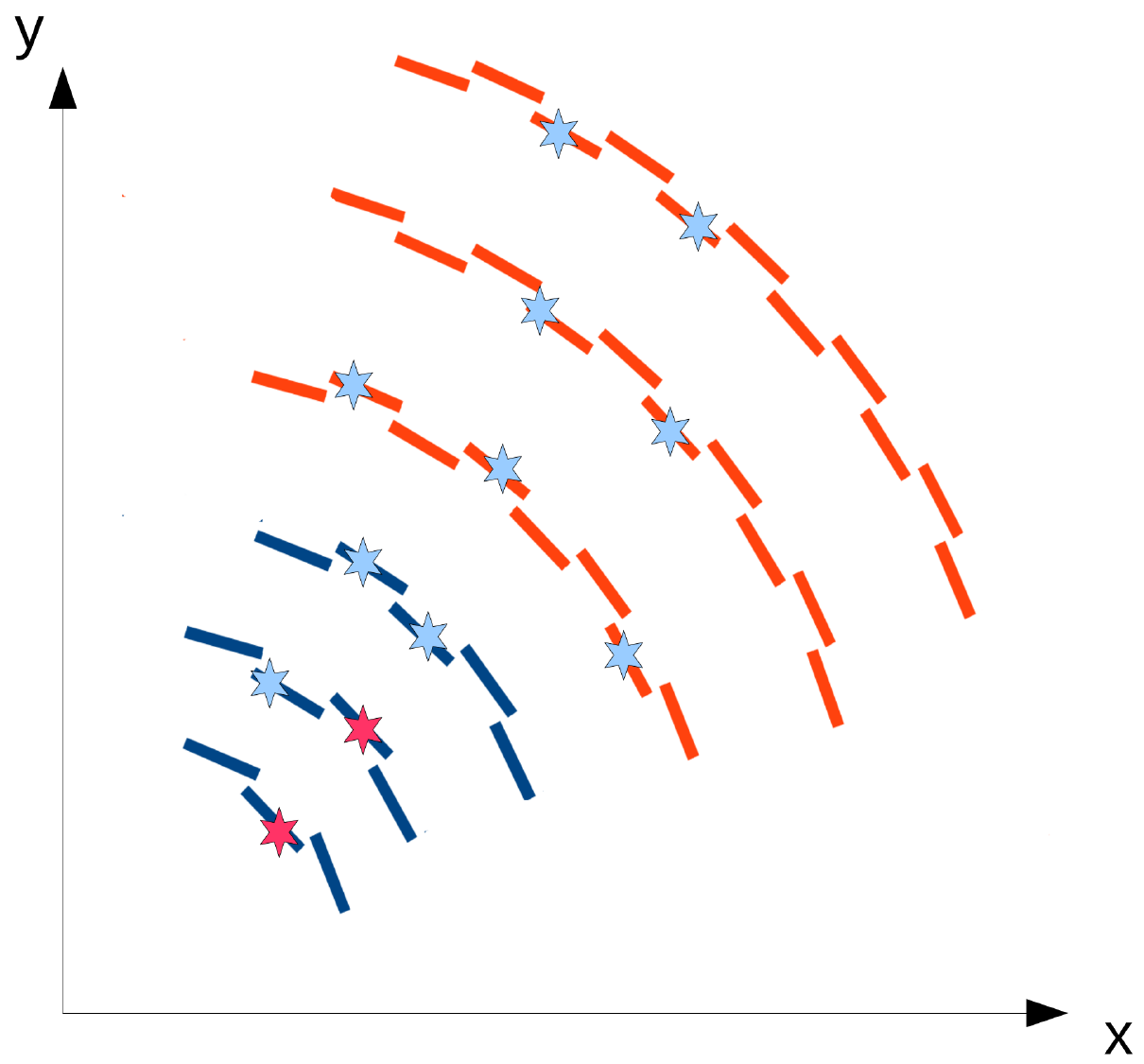}
\includegraphics[width=0.3\linewidth]{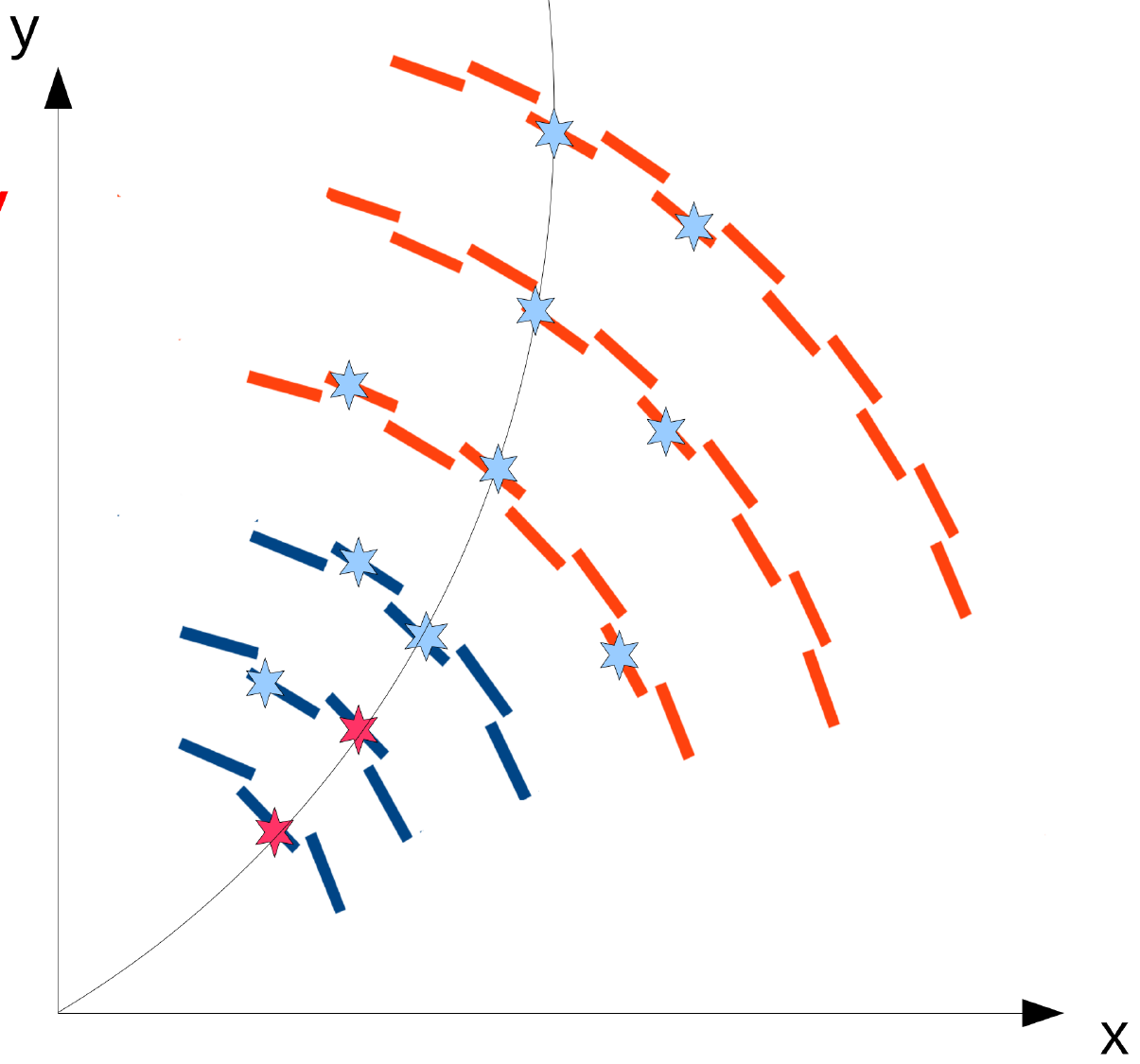}
\includegraphics[width=0.3\linewidth]{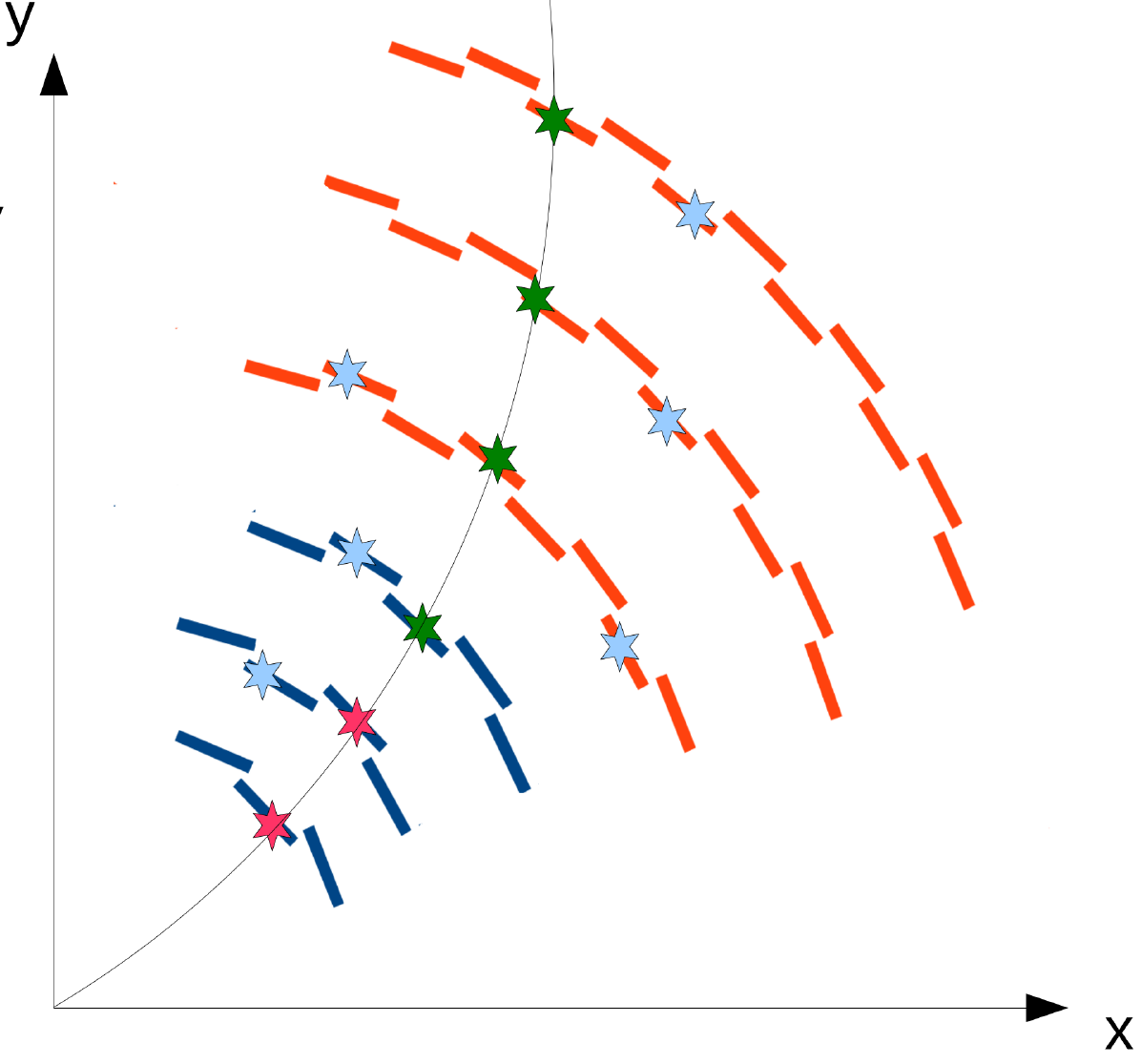}
\caption{Left: the seeding step of the algorithm, in which a tracklet is formed from a pair of stubs (red) in adjacent layers/disks. 
Middle: the projection step, in which the trajectory estimated from the tracklet is projected to other layers/disks. 
Right: the matching and track fitting steps, in which stubs in other layers/disks (green) that are consistent with the projected trajectory are associated to the tracklet (red), and the final track parameters are computed based on both the seed and the matched stubs.}
\label{fig:tracklet_algo}
\end{center}
\end{figure*}

A brief overview of the Tracklet algorithm is shown in Figure~\ref{fig:tracklet_algo}. The Tracklet approach is a road search algorithm similar to what is used in the offline reconstruction. The algorithm makes use of the full precision in the pattern recognition, and is based on massive parallelization achieved via partitioning and linking the detector sub-regions. With the Tracklet approach, the tracker is divided into 27 (current baseline, but configurable) sectors in azimuthal angle ($\phi$), each of which is processed by one FPGA. Any track with $\pt > 2~\GeV$ can only span at most two sectors. 
%

The algorithm starts with forming seeds (i.e. tracklets) between two adjacent layers/disks. 
For parallel processing, each seeding layer/disk of a $\phi$ sector is further divided into virtual modules (VM): 24 $\phi$ divisions for inner layers/disks of the seeding pair, and 16 $\phi$ divisions for the outer ones, which in addition are split into $8~z$ bins. 
A tracklet is built from one stub in the inner VM and one in the outer VM. Each pair of inner and outer VMs are processed in parallel. 
With the above VM configuration, there are in total $24 \times 16 = 384$ pairs of VMs to be processed for a given seeding layer/disk pair. However, only about 120 pairs of the combinations are consistent with $\pt > 2~\GeV$ tracks, and thus only these pairs are connected in the system configuration. Doing so gives us a large reduction of combinatorics, thus resources needed for forming tracklets. 
In order to achieve a good coverage of the entire pseudo-rapidity ($\eta$) range of the tracker, the seedings are done in multiple layer/disk pairs in parallel, including pairs between layer 1 and 2, 3 and 4, 5 and 6, disk 1 and 2, disk 3 and 4, layer 1 and disk 1, layer 2 and disk 1 in the current implementation. 

With a tracklet, initial track parameters are computed assuming the track originated from the beamline. The projected positions to other layers/disks from this tracklet are also calculated and are routed into the corresponding virtual modules based on the projected $\phi$ coordinates. 
The closest stub in the VM to the projection is then associated to the seed, and the differences in $\phi$ and $z$ (or $r$) coordinates between the matched stub and the projection are calculated. 
A linearized $\chi^2$ fit is performed to update the initial tracklet parameters based on the additional information from the residuals of the matched stubs, and compute the final track parameters \pt, $\eta$, $\phi_0$, $z_0$ (and optionally $d_0$). To implement the track fit on the FPGA, the complex calculations involving derivatives used in the fit are pre-computed and tabulated in the lookup tables.

Mostly due to seeding in multiple pairs of layers/disks, one track can be reconstructed more than once. 
Tracks are compared in pairs and are tagged as duplicates by counting the number of independent and shared stubs. The one with higher $\chi^2$ per degree of freedom is removed. 

The above algorithm is implemented via multiple processing steps: 
the ``Layer Router'' and ``VM Router'' steps are for stub organization; 
``Tracklet Engine'' and ``Tracklet Calculator'' are used to form tracklet seeds and calculate initial track parameters. 
``Projection Transceiver'' and ``Projection Router'' steps are for routing projections and transmitting them to other sectors if necessary. 
Matched stubs are found by ``Match Engine''. 
The residuals are computed by ``Match Calculator'' and are transmitted in ``Match Transceiver'', if necessary, to where the projection was produced. 
The track fit is done in ``Track Fit'', 
 and duplicates are removed in the ``Duplicate Removal'' step.

\subsection{TMTT algorithm}
\label{sec:tmtt_algo}
\begin{figure*}
\begin{center}
\includegraphics[width=0.9\linewidth]{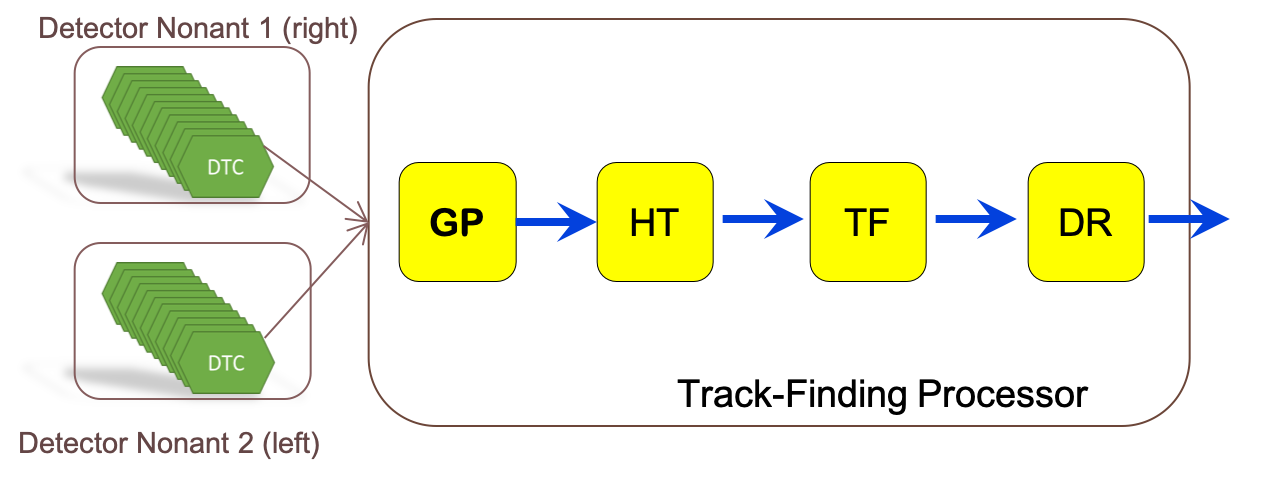}
\caption{An overview of the four main steps of the TMTT algorithm. GP: Geometric Processor for data organization; HT: Hough Transform for pattern recogntion; TF: Track Fitter for filtering tracks and fitting track parameters; DR: Duplicate Removal for removing duplicate tracks.}
\label{fig:tmtt_algo}
\end{center}
\end{figure*}

The core parts of the Time-Multiplexed Track Trigger (TMTT) approach are a Hough Transform for track finding, and a Kalman Filter for the track fit. The algorithm can be grouped into four self-contained logical blocks: Geometric Processor, Hough Transform, Track Fitter and Duplicate Removal, as shown in Figure~\ref{fig:tmtt_algo}. 
%
In the TMTT approach, the tracker and the input stubs are partitioned into $\phi$ nonants. 
Each Geometric Processor (GP) receives stubs from two adjacent detector nonants, and further divides its processing nonant into $2~\phi\times18~\eta$ sub-regions. The track finding in each sub-region occurs in parallel. 
The GP routes and duplicates stubs if they are consistent with more than one sub-region due to track curvature. The stub data are also formatted in GP for more convenient use downstream.

The Hough Transform (HT)~\cite{hough_transf_1, hough_transf_2}, a widely used technique for feature extraction, is used to search for primary tracks in the $r-\phi$ plane. 
The track trajectory in the $r-\phi$ plane is, to a good approximation, a circle within the tracker volume. 
For a given stub coordinate $(r, \varphi)$, assuming the track originated from the beamline, track parameters \pt and $\phi$ to first order must satisfy $\phi = \varphi + r \times q/\pt$, where $q$ is the charge of the particle.
This equation indicates that one stub with coordinates $(r, \varphi)$ maps into a straight line in the track parameter space $(q/\pt, \phi)$, known as Hough space. 
Stubs from the same track will go through a single point in the Hough space, thus the intersection of stub lines can be used to find track candidates.
To implement this on FPGAs, the Hough space is divided into $32\times 64$ cells. The HT computes an array of $(q/\pt, \phi)$ for each stub and fills the corresponding cells. 
A cell that is consistent with 4 or 5 stubs is selected as a track candidate.

The track candidates found by the HT are passed to the Track Fit step using a Kalman filter to fit track parameters.
The fit starts from the coarse helix parameters identified by the HT. Stubs associated to the track candidate are processed one by one, and the track parameters are updated from the stubs iteratively, weighted by their relative uncertainties.
Stubs that are inconsistent with the extrapolation are skipped.

Due to the finite cell size in the implementation of the HT, a single track can be consistent with multiple cells in the Hough space. As a result, over half of the track candidates found by the HT are duplicates. Removal of these duplicated tracks is done by selecting only the cell with parameters that are consistent with the fitted track parameters. This duplicate removal algorithm only looks at individual tracks, and no pairwise comparison between tracks is needed.

\section{Hardware Demonstrators}
\label{sec:demo}
Both the Tracklet and TMTT algorithms are implemented on the hardware demonstrators. The goal of the demonstrators is to show the feasibility and performance of doing L1 track finding with an all-FPGA system.
Both demonstrators of Tracklet and TMTT approaches cover one time-multiplexing slice and one partition of the detector regions. The rest of the entire system would be duplicates of the demonstrators. 
Both systems are built on $\mu$TCA boards with the Xilinx Virtex-7 (XC7VX690T) FPGA~\cite{virtex7}. 
The final L1 track finding system is foreseen to be built on ATCA blades with the more powerful Virtex Ultrascale+ FPGA.
\subsection{Tracklet demonstrator}
\label{sec:tracklet_demo}
The Tracklet demonstrator presented in this proceeding adopts a time-multiplexing factor of 6, so the system processes a new event every $6 \times 25~\mathrm{ns} = 150~\mathrm{ns}$. 
Spatially, the demonstrator covers half of the $z$ region ($z > 0$) and 3~$\phi$ sectors out of total 27 sectors.
Two complete implementations are used to demonstrate the feasibility to cover full $\eta$ range of the tracker: one for a half barrel focusing on seedings with barrel layer pairs; the other for a quarter barrel plus the forward endcaps focusing on seedings with layer-disk pairs and disk-disk pairs. The implementations discussed in this proceeding adopt an earlier version of the VM configuration. Tracking results of both implementations in hardware are compared with the results from integer emulation in software. Excellent agreement in the track parameters are achieved between hardware and software emulation with single muon samples and also top quark pair ($\ttbar$) events with an average PU of 200.

\begin{figure*}
\begin{center}
\includegraphics[width=0.69\linewidth]{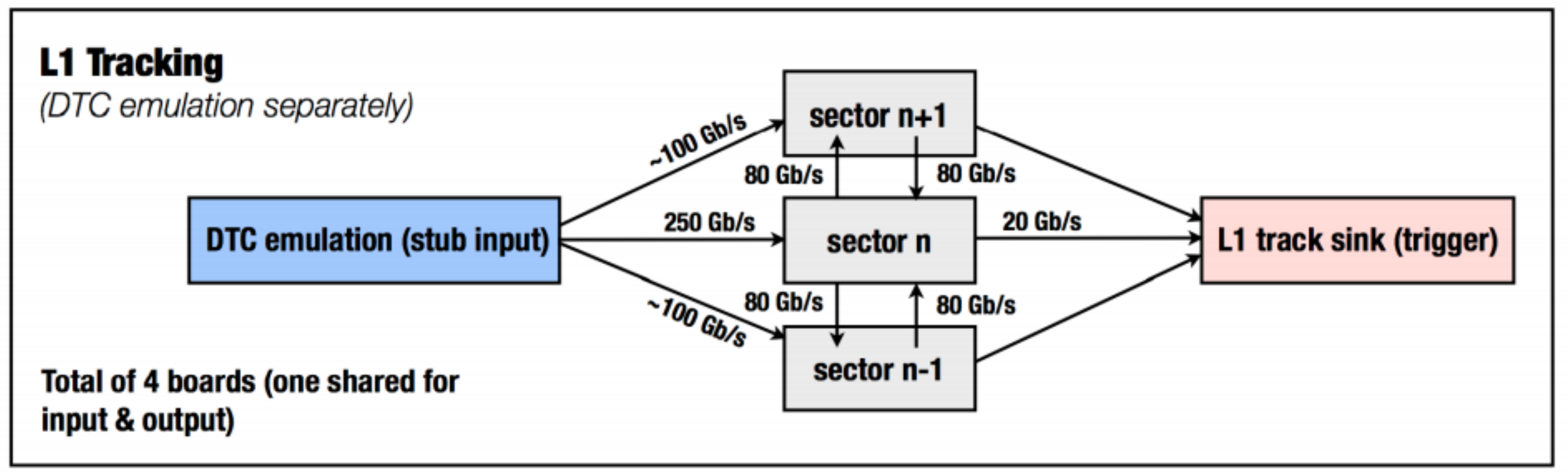}
\includegraphics[width=0.25\linewidth]{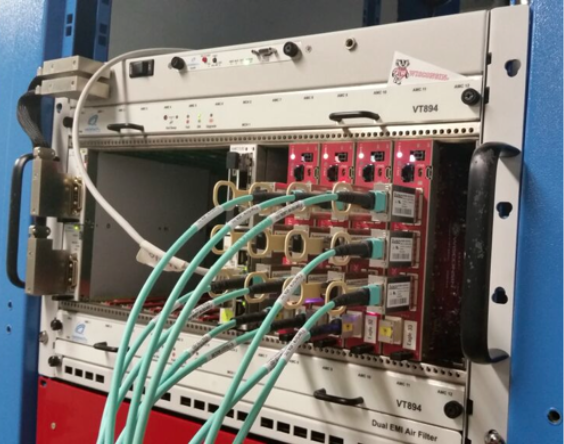}
\caption{A schematic overview (left) and a photo (right) of the Tracklet demonstrator system. 
The input stub source and track sink are on the same CTP7 board. 
Estimated data transfer rates on the inter-board links are also shown.}
\label{fig:tracklet_demo}
\end{center}
\end{figure*}

The boards used for the Tracklet demonstrator are the same boards used in the current CMS L1 trigger, named CTP7~\cite{ctp7}. A CTP7 board hosts one Xilinx Virtex-7 FPGA and a Xilinx Zynq processor for control and communication. 
A total of four CTP7 boards are used, as shown in Figure~\ref{fig:tracklet_demo}: one for the central $\phi$ sector, two for its $\pm \phi$ neighboring sectors and one for sending input stub data as well as collecting the output tracks. 
An AMC13~\cite{amc13} card distributes a central 240 MHz clock signal to all the four CTP7 boards. 
The inter-board communication uses 10 Gb/s links with 8b/10b encoding.
%

Each step in the Tracklet algorithm takes a fixed number of clock cycles to process the input data and write to its output memories. The latency of each processing module consists of the time needed (1-50 clock cycles) to get the first results, plus the total processing time it has (150 ns for a time-multiplexing factor of 6 with 40 MHz input data rate) before moving to the next event. Table~\ref{tab:tracklet_latency} shows the calculated latency of each processing step as well as of the total system, assuming the clock frequency is 240 MHz. 
For some of the steps, data have to be transmitted to the neighboring sector boards. In this case, the latency due to inter-board communication is also included. 
The total estimated latency for receiving the first track from an event is 3345.8 ns. The final processed track for this event should come within 150 ns after the first track due to the fixed processing time for each step.
\begin{table}
  \centering 
  \begin{tabular}{|lrrrrr|}
    \hline 
    Step & Proc.  & Step         & Step  & Link  & Step  \\
         & time        & latency      & latency & delay  & total \\ 
         & (ns)        & (CLK)        & (ns)  &  (ns) & (ns) \\ \hline 
    Input link              &    0.0 &    1  &   4.2   &  316.7  &  320.8   \\  
    Layer Router            &  150.0 &    1  &    4.2   &   -  &  154.2   \\  
    VM Router               &  150.0 &    4  &   16.7   &   -  &  166.7        \\  
    Tracklet Engine         &  150.0 &    5  &   20.8   &   -  &  170.8        \\  
    Tracklet Calculation    &  150.0 &   43  &  179.2   &   -  &  329.2        \\  
    Projection Trans.  &  150.0 &   13  &  54.2   & 316.7  &  520.8      \\  
    Projection Router       &  150.0 &    5  &   20.8   &  -  &  170.8        \\  
    Match Engine            &  150.0 &    6  &   25.0   &   -  &  175.0        \\  
    Match Calculator        &  150.0 &   16  &   66.7   &   -  &  216.7        \\  
    Match Trans.       &  150.0 &   12  &  50.0   & 316.7  &  516.7      \\  
    Track Fit                    &  150.0 &   26  &  108.3   &   -  &  258.3        \\  
    Duplicate Removal     &  0.0 &   6  &   25.0   &   -  &  25.0        \\  
    Output Link             &    0.0 &    1  &    4.2   & 316.7  &  320.8        \\  
\hline 
    Total                   & 1500.0 &  139  &  579.2   & 1266.7  & 3345.8    \\
    \hline
  \end{tabular}
  \caption{Tracklet demonstrator latency model. The processing time and latency for each step as well as for the total chain are given. For steps involving data transfers, the link latency is given.}
  \label{tab:tracklet_latency}
\end{table}
The total latency of the algorithm is also measured with a clock counter on the demonstrator. The measured latency for receiving the first track is 800 clock cycles with a 240 MHz clock, or 3333~ns. 
The measured latency agrees within 3 clock cycles (0.4\%) with the estimated latency, both of which meet the 4 \mus latency goal for L1 track finding.
In an improved configuration of the project, the data transmission between processing sectors is no longer needed, at a cost of duplicating some stubs at the boundary of sectors. In that case, the total latency could be further reduced.

\subsection{TMTT demonstrator}
\label{sec:tmtt_demo}
\begin{figure*}
\begin{center}
\includegraphics[width=0.8\linewidth]{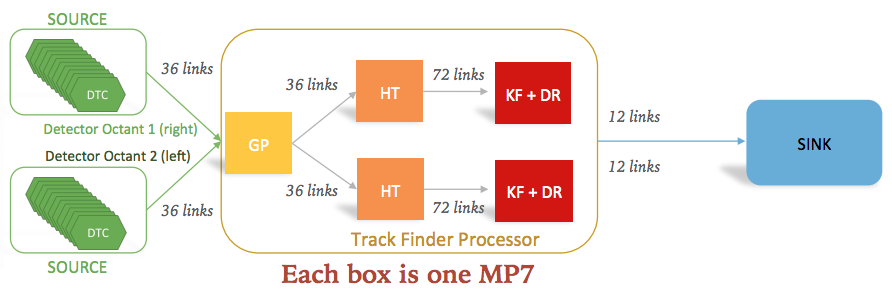}
\includegraphics[width=0.5\linewidth]{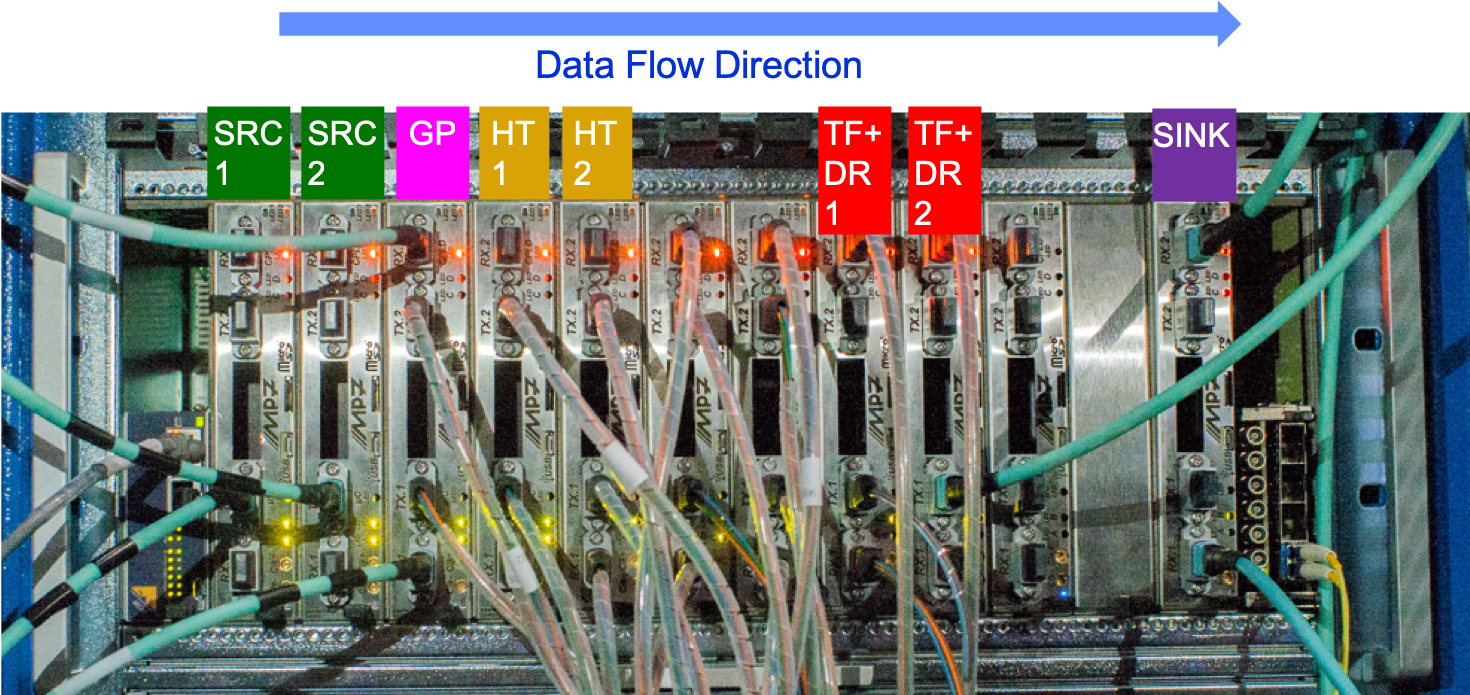}
\caption{A schematic overview (top) and a photo (bottom) of the TMTT demonstrator system. 
The number of optical links used between the logical blocks are shown.}
\label{fig:tmtt_demo}
\end{center}
\end{figure*}

The TMTT demonstrator shown in this proceeding is based on a time-multiplexing factor of 36, reconstructing tracks in $\phi$ nonants as discussed in Sect.~\ref{sec:tmtt_algo}. 
Figure~\ref{fig:tmtt_demo} shows an overview of the TMTT demonstrator system.
The TMTT demonstrator is implemented on the Imperial Master Processor with Virtex-7 FPGA (MP7)~\cite{mp7}. 
Each box in Figure~\ref{fig:tmtt_demo} corresponds to one MP7 board. Eight MP7 boards are used for the demonstrator chain. 
Two boards are used as stub data source, each of which represents a detector nonant. The stub data are sent into the boards via IPBus. 
The output data streams of the two source boards are sent to one MP7 board functioning as the Geometric Processor. 
Two boards are used for the Hough Transform, two more are used for Kalman Filter plus Duplicate Removal.
One additional board is used as the track sink collecting tracks from the track finder, which are then read out via IPBus. 
With the demonstrator setup, the hardware output can be directly compared with software emulation.

\begin{table}
  \centering
  \begin{tabular}{|lr|}
    \hline
    System & Latency (ns)  \\ 
    \hline 
    SERDES + optical length & 143 \\
    Geometric Processor & 251 \\
    SERDES + optical length 2 & 144 \\
    Hough Transform & 1025 \\
    SERDES + optical length 3 & 129 \\
    Kalman Filter + Duplicate Removal & 1658 \\
    SERDES + optical length 4 & 129 \\
    \hline
    Total: First out - First in & 3538 \\
    Last out - First out & 225 \\
    Total: Last out - First in & 3763 \\
    \hline
  \end{tabular}
  \caption{TMTT demonstrator latency model. Latencies from processing steps as well as from optical links are shown. The time difference between when the first stub arrives at the track finder and when the last track is sent out is shown, in addition to the time difference between when the first stub is received and when the first track is sent out.}
  \label{tab:tmtt_latency}
\end{table}
The latency of the TMTT demonstrator is fixed, regardless of the pileup and event occupancy in the nonant.
The latency is measured both for each step independently and for the entire chain. The sum of the individual measurements gives identical result compared to the measurement on the full chain from source to track sink. 
Both the processing blocks and the optical links contribute to the total latency. The link latency includes optical transmission delays and serialization/deserialization (SERDES) latency.
The results are summarized in Table~\ref{tab:tmtt_latency}. The total latency between when the first stub arrived at the track finder and when the first track is sent out is 3538~ns, which is below the 4 \mus latency budget for L1 track finding. An additional 225 ns is expected for reading out the last track in this event. 

\section{Performance Studies}
\label{sec:performance}

The demonstrators of both Tracklet and TMTT described in Sect.~\ref{sec:demo} showed good tracking performance. The Tracklet performance is discussed in Sect.~\ref{sec:tracklet_perf}, followed by the discussion of TMTT performance in Sect.~\ref{sec:tmtt_perf}.

\subsection{Tracklet performance}
\label{sec:tracklet_perf}
\begin{figure*}
\begin{center}
\includegraphics[width=0.50\linewidth]{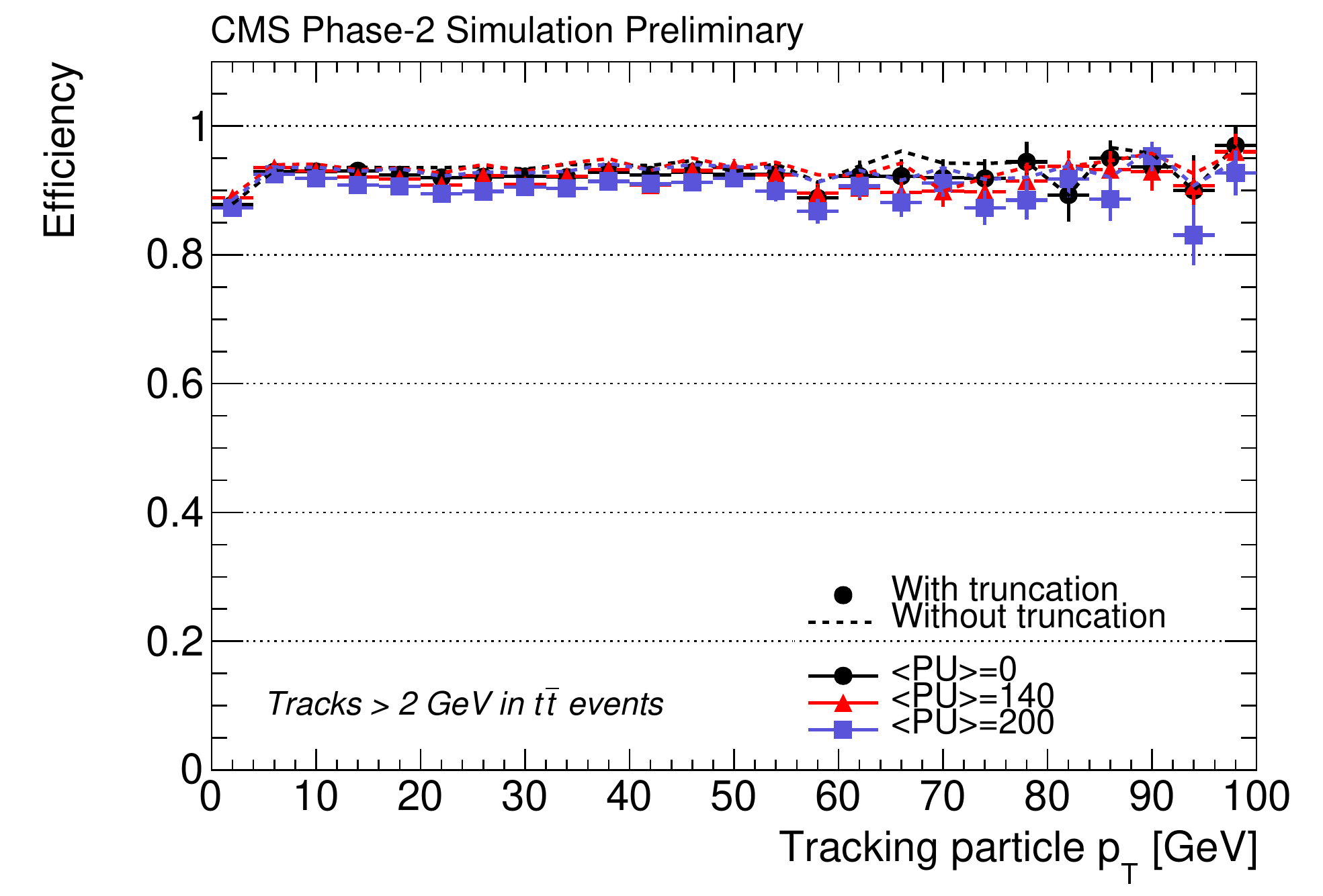}
\caption{Efficiencies as a function of particle \pt for $\ttbar$ events. Results are shown for three average pileup scenarios: 0 (black), 140 (red) and 200 (blue). The efficiencies with (solid marker) and without (dashed lines) truncation effects are shown.}
\label{fig:tracklet_eff}
\end{center}
\end{figure*}

The perfect agreement in the reconstructed track parameters between the integer-based C++ emulation and the hardware demonstrator let us safely study the performance of the L1 track finding system in software emulation.
The tracking efficiencies as a function of tracking particle \pt in simulated $\ttbar$ events with average PU of 0, 140 and 200 are shown in Figure~\ref{fig:tracklet_eff}. Both efficiencies with and without truncating data due to fixed latency are shown in the plot. Overall quite good efficiencies (> 95\%) are achieved even in the very busy $\ttbar$ events. 

\begin{figure*}
\begin{center}
\includegraphics[width=0.49\linewidth]{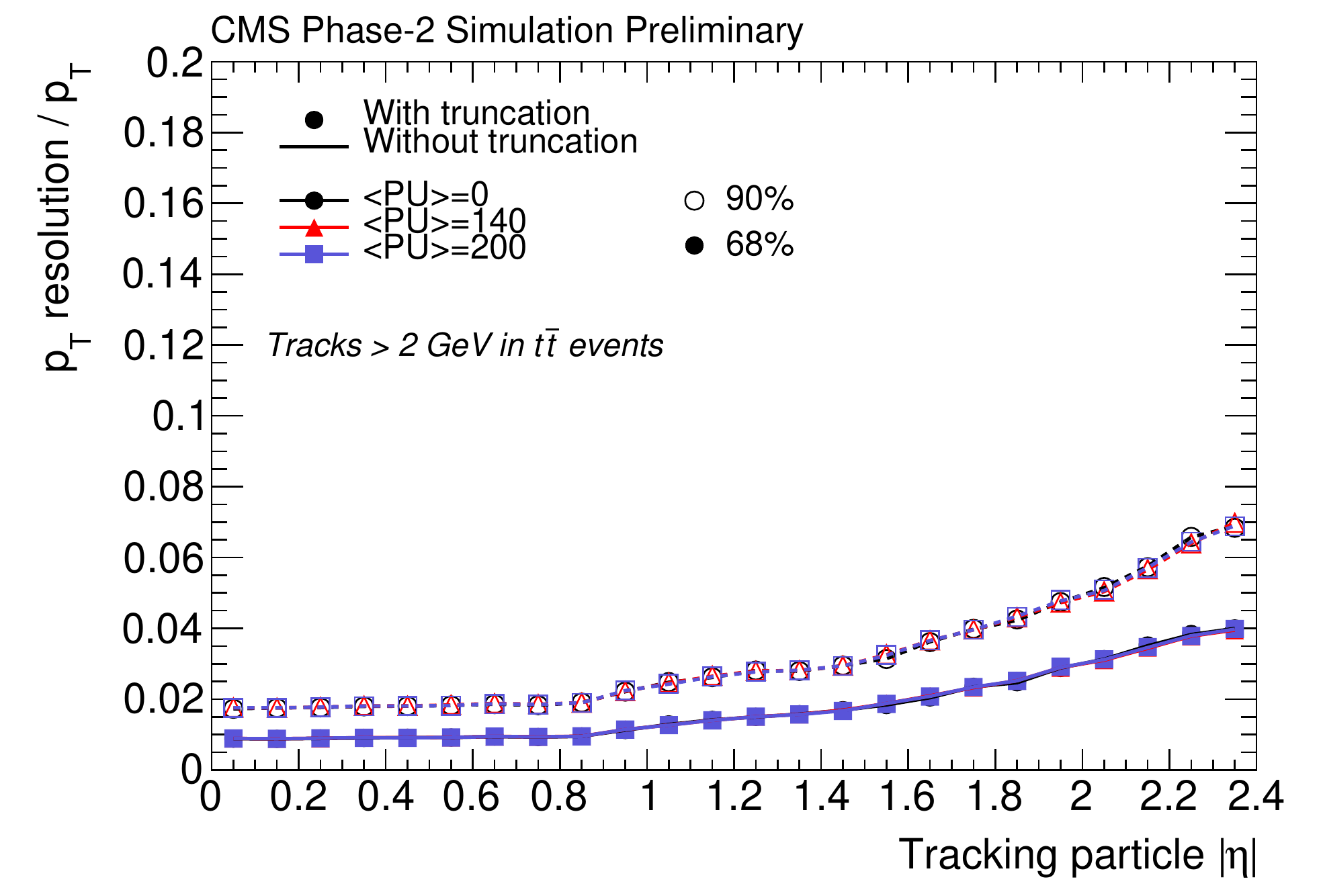}
\includegraphics[width=0.49\linewidth]{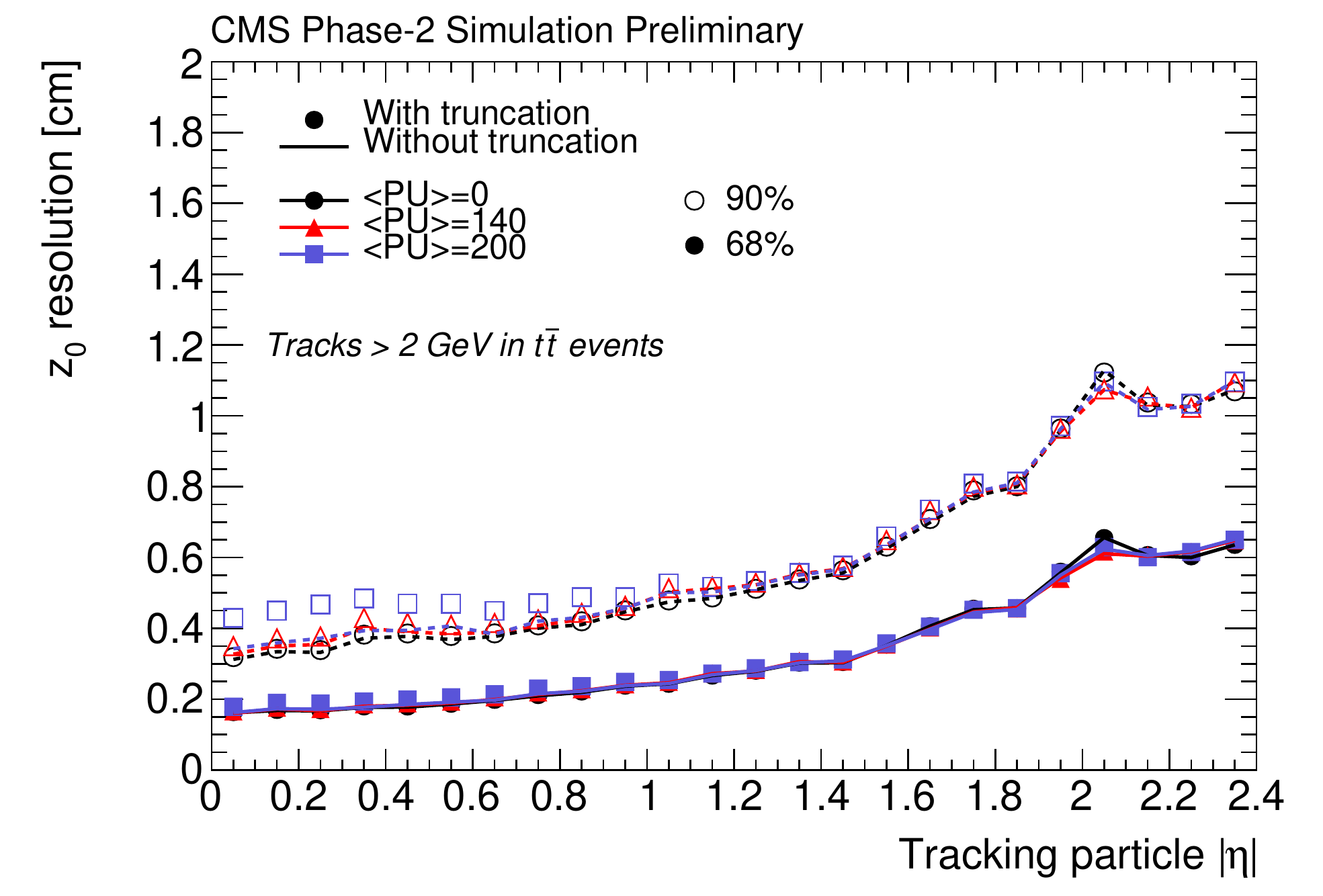}
\caption{Resolutions of relative \pt (left) and $z_0$ (right) in the final track parameters for $\ttbar$ events with 0 (black), 140 (red) and 200 (blue) pileup events on average. 
The resolutions shown in the plots correspond to intervals that encompass 68\% (filled markers and solid lines) or 90\% (open markers and dashed lines) of the tracks in the \pt or $z_0$ distributions with (markers) or without (lines) truncation effect.}
\label{fig:tracklet_res}
\end{center}
\end{figure*}

The resolutions of track parameters \pt and $z_0$ as a function of tracking particle $|\eta|$ are shown in Figure~\ref{fig:tracklet_res} for $\ttbar$ events with 0, 140 and 200 pileup events on average. 
The plots show that the tracklet algorithm achieves the resolution required of the L1 trigger system, specifically a $z_0$ resolution of 2-3 mm in the barrel, and a relative \pt resolution below 4\% for the entire $\eta$ coverage.

\subsection{TMTT performance}
\label{sec:tmtt_perf}
\begin{figure*}
\begin{center}
\includegraphics[width=0.49\linewidth]{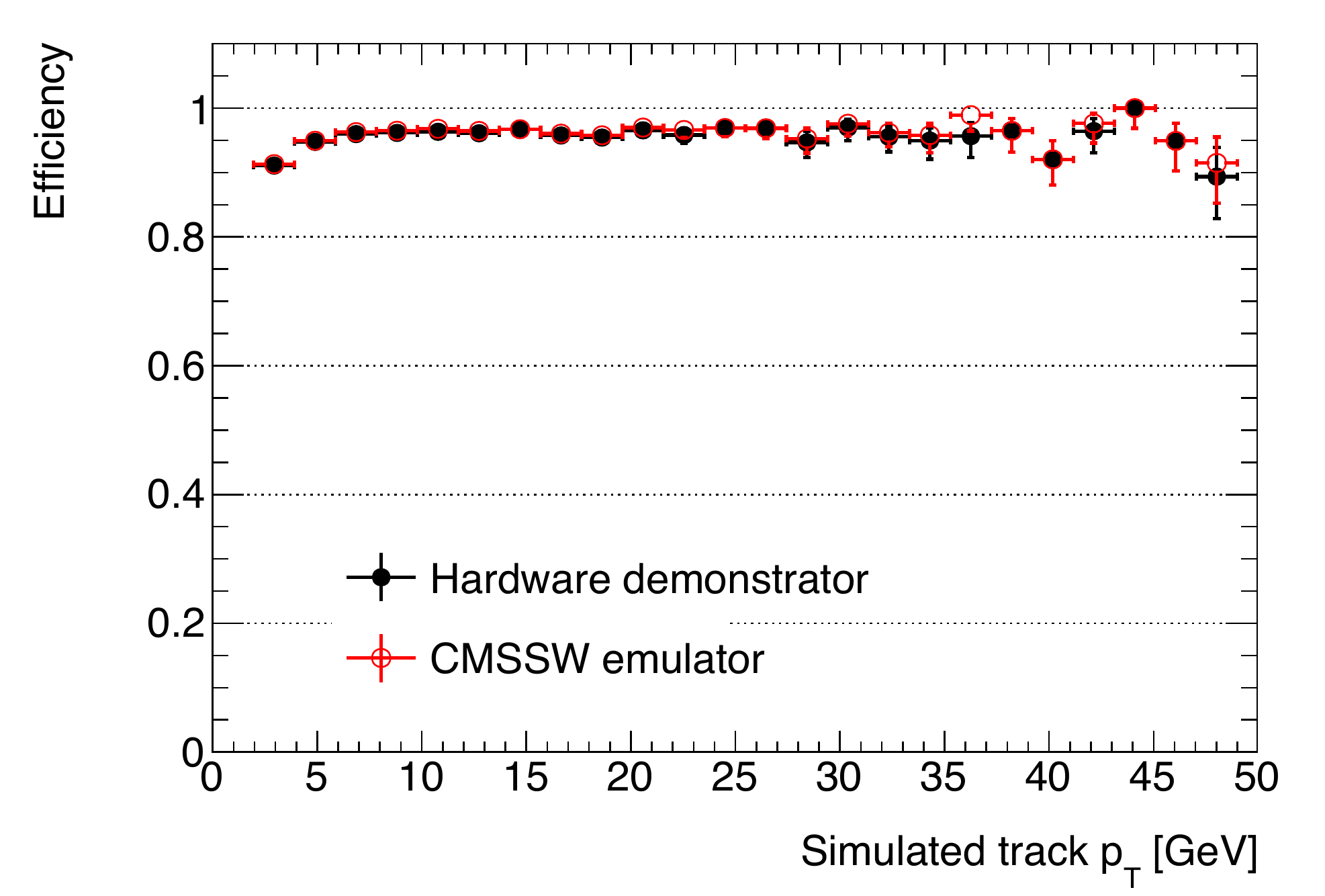}
\includegraphics[width=0.49\linewidth]{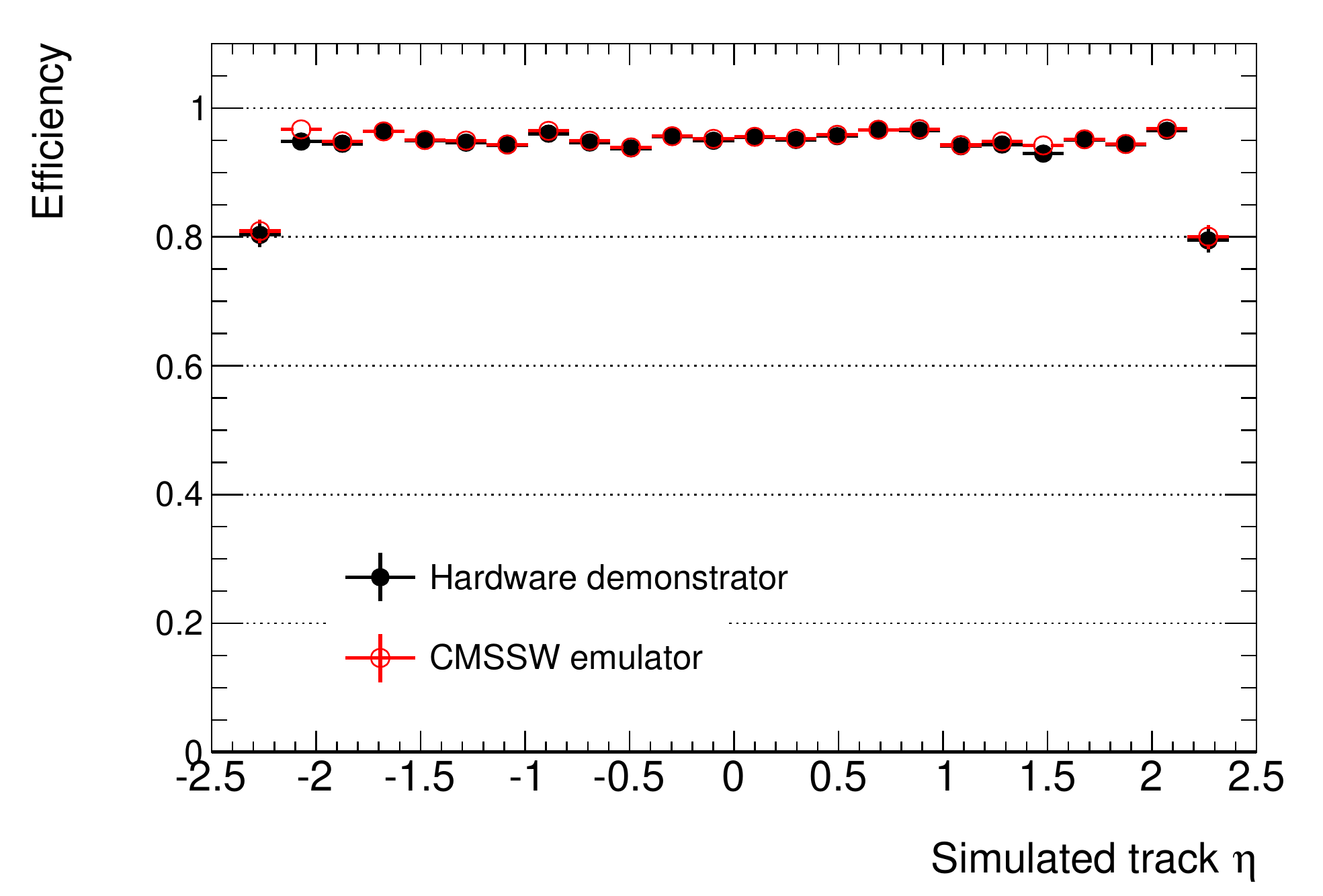}
\caption{Tracking efficiencies as a function of particle \pt (left) and $\eta$ (right) with $\ttbar$ events plus 200 pileup events. Both results from hardware demonstrator (black) and software emulation (red) are shown.}
\label{fig:tmtt_eff}
\end{center}
\end{figure*}

With the setup of the TMTT demonstrator discussed in Sect.~\ref{sec:tmtt_demo}, the results as well as the tracking performance from the hardware demonstrator are directly compared with the same events processed in the software emulation. 
Figure~\ref{fig:tmtt_eff} shows the track finding efficiencies for tracks with $\pt > 3~\GeV$ as a function of tracking particle \pt or $\eta$ in simulated $\ttbar$ events with 200 PU.
Greater than 95\% tracking efficiencies are achieved in most of the \pt and $\eta$ range. 
The tracking efficiencies are also in good agreement between the hardware demonstrator and software emulation.

\begin{figure*}
\begin{center}
\includegraphics[width=0.47\linewidth]{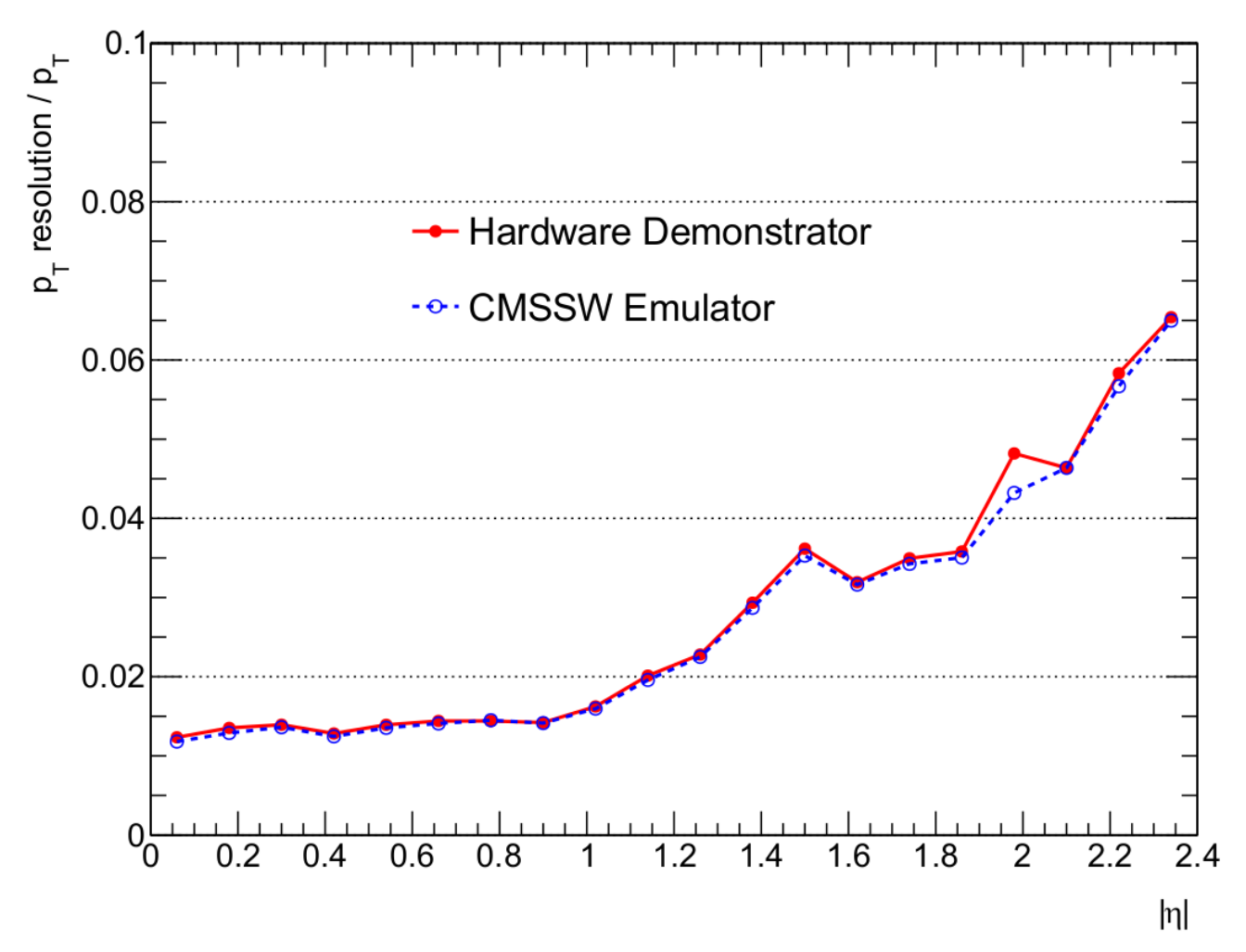}
\includegraphics[width=0.47\linewidth]{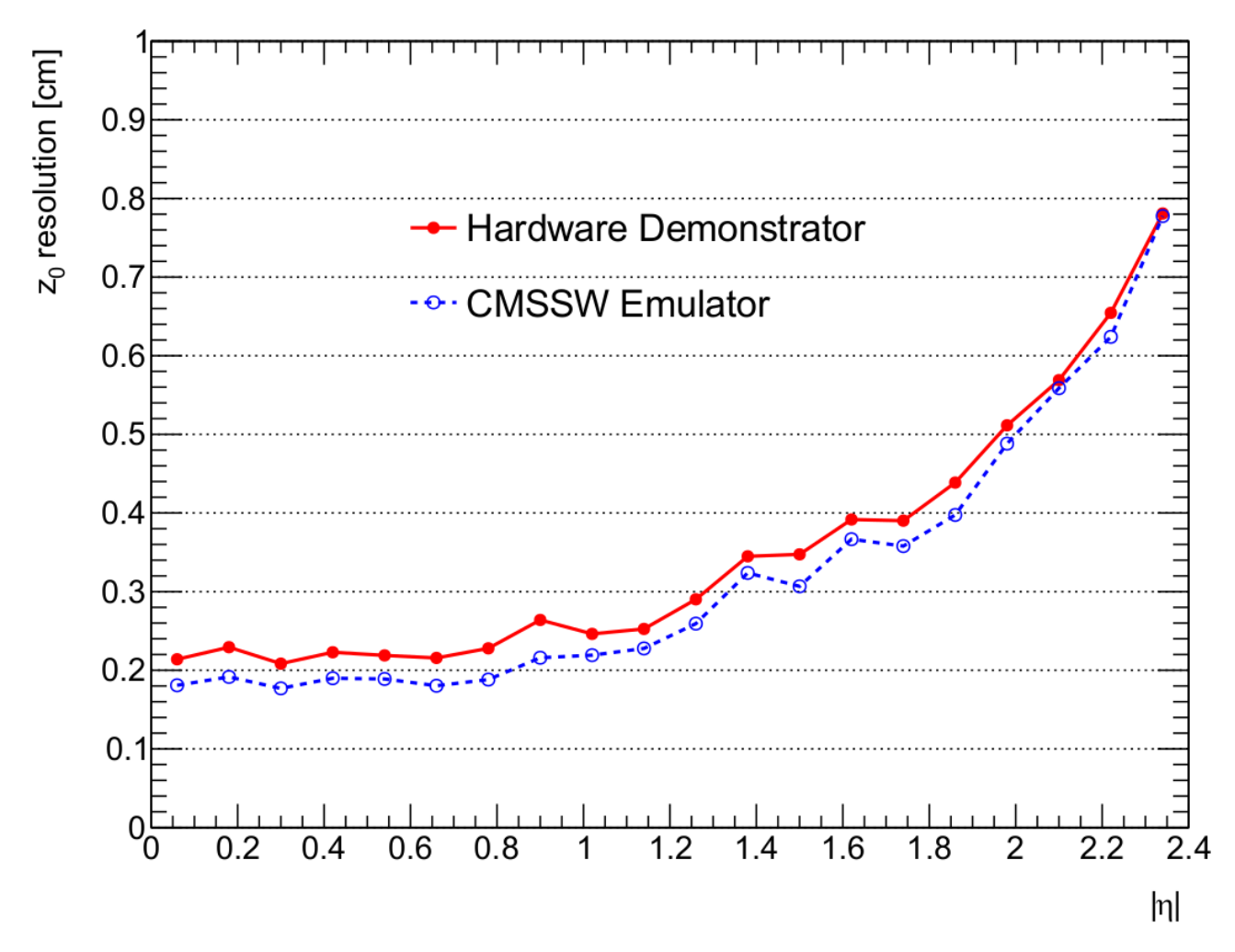}
\caption{Resolutions of relative \pt (left) and $z_0$ (right) as a function of $|\eta|$ in $\ttbar$ events plus 200 pileup events. Results from hardware demonstrator (red) and software emulation (blue) are shown.}
\label{fig:tmtt_res}
\end{center}
\end{figure*}

The resolutions of the final track parameters \pt and $z_0$ are shown as a function of $|\eta|$ in Figure~\ref{fig:tmtt_res}.
Good resolutions of the track parameters for the L1 trigger as well as good agreement between the hardware demonstrator and the software emulation are achieved.

\section{Conclusion}
A Level-1 track trigger at the HL-LHC is necessary in order to maintain high trigger efficiency and manageable rates. The new design of the CMS tracker for the HL-LHC era is largely driven by the requirement to provide tracking information to the L1 trigger. 
Two all-FPGA approaches, Tracklet and TMTT, to implement L1 track finding are presented. Both approaches are based on highly parallelized tracking algorithms, and have been successfully implemented on hardware demonstrators, showing both the feasibility and good performance. 
Effort has been started to bring the two approaches together, including developing a common hardware infrastructure and defining a reference algorithm combining the strengths of the two.

\section*{Acknowledgements}
This work is supported by the US National Science Foundation through the grant NSF-PHY-1307256. 
I would also like to thank the CMS collaboration and my colleagues involved in both Tracklet and TMTT development.

 \bibliography{ref.bib}

\end{document}